# *Ab initio* screening of metallic MAX ceramics for advanced interconnect applications

Kiroubanand Sankaran[1], Kristof Moors[2], Zsolt Tőkei[1], Christoph Adelmann[1], and Geoffrey Pourtois[1,3]

[1]Imec, Kapeldreef 75, 3001 Leuven, Belgium

[2]Forschungszentrum Jülich, Peter Grünberg Institut, 52425 Jülich, Germany

[3]University of Antwerp, Department of Chemistry, 2610 Wilrijk-Antwerpen, Belgium

The potential of a wide range of layered ternary carbide and nitride MAX phases as conductors in interconnect metal lines in advanced CMOS technology nodes has been evaluated using automated first-principles simulations based on density functional theory. The resistivity scaling potential of these compounds, *i.e.* the expected sensitivity of their resistivity to reduced line dimensions, has been benchmarked against Cu and Ru by evaluating their transport properties within a semiclassical transport formalism. In addition, their cohesive energy has been assessed as a proxy for the resistance against electromigration and the need for diffusion barriers. The results indicate that numerous MAX phases show promise as conductors in interconnects of advanced CMOS technology nodes.

## I. INTRODUCTION

The perennial downscaling of electronic circuits in complementary metal-oxide-semiconductor (CMOS) technology has led to a gradual deterioration of the resistance and reliability of the interconnects with the consequence that interconnects today strongly affect the performance and reliability of the overall circuit [1–6]. For over two decades, Cu has been used as the main conductor material and soon, interconnect lines will approach widths of 10 nm or below. At such small dimensions, the resistivity of Cu increases strongly due to surface and grain boundary scattering [7–12]. This can be attributed to the long intrinsic mean free path (MFP) of the charge carriers in Cu (about 40 nm) since the resistivity increase at small dimensions scales approximately with $r_0 \times {}^{l}/_{d}$, where $\rho_0$ is the bulk resistivity, $\lambda$ is the MFP, and $d$ a characteristic dimension (film thickness for surface scattering or grain size for grain boundary scattering) [7,13]. Hence, metals with a shorter MFP should intrinsically be less susceptible to surface or grain boundary scattering for a given $d$ [8,14,15] and may outperform Cu at small dimensions despite their larger bulk resistivity. This has recently led to the concerted search for alternative metals with short MFPs as potential Cu replacements [2,4,8,16–21].

Moreover, Cu interconnects require diffusion barriers and adhesion liners to ensure their reliability, since dielectric breakdown and electromigration (EM) become increasingly problematic at small dimensions [4,22]. Yet, the thickness of barrier and liner layers (with typically high resistivity) does not scale well with the interconnect line dimension and thus they occupy an increasing part of the metal volume without contributing much to the conductance. By contrast, metals with a higher cohesive energy than Cu can be expected to be more resistant to EM and diffusion into dielectrics [23]. It is therefore crucial to identify alternatives to Cu that display both low resistivity and an improved EM performance in ultrasmall interconnect lines without the need for barrier or liner layers.

In this context, it is important to rank the potential of alternative conductor materials in terms of their intrinsic resistivity, MFP, and EM resistance. Among the myriad of possible metallic materials, elemental metals are naturally the simplest conductors and have therefore elicited much of the initial interest. Among them, Co, Mo, and several Pt-group metals have recently emerged as alternatives due to their combination of low bulk resistivity, short MFP, high melting point, and good EM performance [14,24]. Lately, Co has been introduced into commercial CMOS circuits albeit still requiring a diffusion barrier to ensure reliability [25,26].

More recently, the screening effort to identify suitable conductor materials has broadened to include also binary and ternary compounds [15,27–30]. Although binary and ternary systems have been studied extensively in the metallurgic community, reports on their thin film resistivity in the relevant thickness range around 10 nm are scarce. In the last decades, a group of metallic hexagonal carbide and nitride ternary ceramics, termed MAX compounds, has prompted much attention for coating applications and as lightweight structural materials owing to the combination of excellent high-temperature mechanical properties, high melting points, large thermal (and thus electrical) conductivity, and good oxidation resistance [27,31–35]. This group of compounds with the generic formula $M_{n+1}AX_n$ consist of an early transition metal (M), an element of columns 13 or 14 of the periodic table (A), and either C or N (X) [31–34]. The compound stoichiometry is described by $n$, which ranges from $n$ = 1 to 3, *i.e.* the stoichiometry can be denoted as 211, 312, or 413. At present, more than 100 different MAX compounds have been identified [33–35] and numerous other combinations are possible.

Interestingly, the above set of material characteristics is also highly relevant for conductors in interconnects. Some MAX compounds have been reported to exhibit bulk in-plane resistivities as low as 10 μΩcm or below [32,32,36], which renders them interesting for scaled interconnect applications if the resistivity shows little sensitivity to the structure or grain size. To further evaluate the potential of MAX phases for advanced interconnects, we have used automated first-principles simulations based on density-functional theory (DFT) to calculate the enthalpy of formation of 170 different MAX compound candidates to establish their stability as well as their cohesive energy as a proxy for resistance to EM. In a second step, the product of the bulk resistivity, $\rho_0$, and the MFP, $\lambda$, has been calculated for each MAX compound as a metric for their resistivity scaling potential in confined dimensions [14]. This benchmarking methodology has been applied to elemental metals before and has led to the identification of several prospective candidates, such as Ru, Ir, Rh, Co, or Mo [14, 17, 24], with subsequent experimental confirmation of their promising properties [24]. Here, we use the same methodology for the identification of potential candidate MAX materials that combine both short MFP and large cohesive energy. While additional experimental work is required to demonstrate the actual suitability and performance of a certain MAX phase in interconnect applications, our approach describes clear directions for the material selection process. Especially, large values of the $\rho_0 \times \lambda$ product of a MAX compound indicate a large value of $\rho_0$ and/or of $\lambda$, which are both detrimental for conductors in scaled interconnects. Thus, both the computed $\rho_0 \times \lambda$ product and the cohesive energy will be used below as figures of merit to rank MAX phases for their potential in interconnect applications.

## II. Methodology

The automated first-principles DFT simulations were based on generic prototypes for the $M_{n+1}AX_n$ crystal structures [31–34] with $n$ = 1 to 3. MAX compounds with 211 composition exist as a single phase, while 312 or 413 MAX compounds can coexist as two polymorphs, termed α and β [31–34], which differ in the sequence and geometry of the stacked hexagonal planes. Representations of the five different crystal structures considered in this paper are shown in Fig. 1. To obtain concrete MAX phases, the different atomic sites were then populated with M ∈ {Ti, V, Cr, Ta}, A ∈ {Al, Si, Ga, Ge, As, Cd, In}, and X ∈ {N, C}. The atomic structure and the lattice parameters of each prototype were then optimized, and the total energy of the structure was used to assess the enthalpy of formation. All MAX phases with negative enthalpy of formation were then selected for further evaluation of their phononic and electronic properties.

The structural optimizations, the evaluation of the formation enthalpy, and the computation of the electronic properties were performed using first-principles DFT simulations, as implemented in the Quantum Espresso package [37]. The first Brillouin zone was sampled using a regular Monkhorst-Pack scheme [38] with a $k$-point density of 60×60×40. The exchange-correlation energy was described within the Perdew–Burke–Ernzerhof generalized gradient approximation [39]. The valence electrons were described using GBRV pseudopotentials [40] with a kinetic energy cutoff between 60 and 80 Ry (depending on the elemental composition) and a Methfessel-Paxton smearing function with a broadening of 13.6 meV. This ensured a convergence of the total energy within $10^{-12}$ eV. Phonon band structure calculations involving the determination of the dynamical matrix were performed with supercell and finite displacement approaches (atomic displacement distance of 0.01 Å), as implemented in the phonopy package [40].

As discussed above, the potential of MAX phases as interconnect conductors can be assessed based on two proxies: the cohesive energy as well as the $\rho_0 \times \lambda$ product [14,15,27]. EM processes in metals, in a first-order approximation, can be described by the intrinsic self-diffusion of metal atoms due to charge currents as well as thermal or mechanical stress gradients. The self-diffusion activation energy is related to the strength of the metallic bonds, *i.e.* the cohesive energy of the metal [23], which is a quantity easily accessible by DFT. For elements, the cohesive energy is approximately linearly correlated to the melting temperature of the metal [23], and therefore the latter has been used for a simple benchmarking of the EM resistance of metals. However, for most MAX phases, the melting temperature is typically not (accurately) known and therefore we have calculated the cohesive energy as a proxy. A second major reliability issue is the diffusion and drift of metal ions into surrounding dielectrics, leading to dielectric breakdown. The dielectric breakdown can be avoided by introducing diffusion barriers for the metal atoms, however with the scaling limitations discussed above. Since it has been found that the limiting mechanism is typically the detachment of metal atoms from an interconnect line rather than the drift or diffusion itself [23], the cohesive energy can also be used as a proxy for the need for diffusion barriers in the interconnect metallization scheme. Hence, the cohesive energy of a metal can be considered as a general proxy for its prospects for interconnect reliability.

Both formation enthalpies and cohesive energies have been derived from the calculated total energies of the compounds and their constituent elements. The formation enthalpy of a $M_{n+1}AX_n$ compound was calculated by

$$\Delta H_\mathrm{f} = E_\mathrm{tot} - (n+1)\mu_M - \mu_A - n\mu_X, \tag{1}$$

with $E_\mathrm{tot}$ the computed total energy of the MAX compound and $\mu_\mathrm{M}$, $\mu_\mathrm{A}$, and $\mu_\mathrm{X}$ the chemical potentials of M, A, and X materials, respectively, which are equal to the DFT total energies of their (crystalline) ground states. By contrast, the cohesive energy was computed according to

$$E_\mathrm{coh} = E_\mathrm{tot} - (n+1)\mu'_M - \mu'_A - n\mu'_X, \tag{2}$$

with $\mu'_M$, $\mu'_A$, and $\mu'_X$ the chemical potentials of *individual* M, A, and X atoms, respectively.

In a second step, the scalability of the resistivity of MAX materials has been assessed. Previously, the product of the bulk resistivity and the MFP of a metal, $\rho_0 \times \lambda$, has been proposed as a figure of merit for its resistivity to small dimensions [14], and has been used to screen elemental metals as well some binary intermetallics [27,28]. In bulk metals, the charge carrier transport can be described by the semiclassical Boltzmann transport equation within the relaxation time approximation. In this framework, the conductivity $\sigma_{\boldsymbol{t}}$ in transport direction $\boldsymbol{t}$ at low temperatures can be written as [24]

$$\sigma_{\boldsymbol{t}} = 2\frac{e^2}{(2\pi)^3}\sum_n \int d^3\boldsymbol{k}\ \left|\boldsymbol{v}_{n,\boldsymbol{t}}(\boldsymbol{k})\right|^2 \tau_n(\boldsymbol{k})\ \delta\big(\epsilon_n(\boldsymbol{k}) - E_{F,n}\big), \tag{3}$$

with $\boldsymbol{v}_{n,\boldsymbol{t}}(\boldsymbol{k})$ the projection of the Fermi velocity $\boldsymbol{v}_n(\boldsymbol{k}) = (1/\hbar)\nabla_{\boldsymbol{k}} E_{F,n}$ on the transport direction, $\tau_n(\boldsymbol{k})$ the relaxation time, $E_{F,n}$ the Fermi energy, and $e$ the elemental charge. Here, the summation occurs over band index $n$ (the factor of 2 accounting for spin degeneracy) and the integration over reciprocal space, *i.e.* over all occupied states with energy $\epsilon_n(\boldsymbol{k})$ and wavevector $\boldsymbol{k}$.

Two approximations have been proposed to evaluate Eq. (3) and gain further insight into the resistivity scaling via the $\rho_0 \times \lambda$ product without the need to calculate the electron-phonon relaxation time $\tau_n(\boldsymbol{k})$, which is computationally expensive [20,24]. Assuming that either the relaxation time $\tau$ or the MFP $\lambda$ is isotropic (constant), leads to an integral that depends only on the morphology of the Fermi surface [20,24]. It has been argued that an assumption of isotropic $\tau$ is preferred in cases when the resistivity is limited by phonon scattering, whereas isotropic $\lambda$ corresponds better to cases of dominant impurity scattering [42]. Nonetheless, the comparison of the two approximations can also serve as a test for the "robustness" of the obtained values under the respective approximations. In this work, we have therefore evaluated the $\rho_0 \times \lambda$ product following both approaches.

Assuming that $\lambda_n(\boldsymbol{k}) = \tau_n(\boldsymbol{k}) \times |v_n(\boldsymbol{k})| \equiv \lambda$, Eq. (3) becomes [20]

$$\text{Constant } \lambda: \quad (\rho_0 \times \lambda)^{-1}_{\boldsymbol{t},\boldsymbol{\lambda}} = \frac{e^2}{4\pi^3 \hbar} \sum_n \iint_{S_F^n} \frac{\left|\boldsymbol{v}^2_{n,\boldsymbol{t}}(\boldsymbol{k})\right|}{|\boldsymbol{v}_n(\boldsymbol{k})|^2} dS, \tag{4}$$

with the integration carried out over the Fermi surface $S_F^n$ of band $n$. Alternatively, assuming a constant isotropic relaxation time $\tau_n(\boldsymbol{k}) \equiv \tau$ leads to [14,24]

$$(\rho_0 \times \tau)_{\boldsymbol{t}}{}^{-1} = \frac{e^2}{4\pi^3 \hbar} \sum_n \iint_{S_F^n} \frac{\left|\boldsymbol{v}^2_{n,\boldsymbol{t}}(\boldsymbol{k})\right|}{|\boldsymbol{v}_n(\boldsymbol{k})|} dS. \tag{5}$$

To convert from relaxation time to MPF, the average MFP $\lambda$ can expressed as $\lambda = \tau \times \langle \boldsymbol{v}_{n,t}(\boldsymbol{k}) \rangle$, with $\langle \boldsymbol{v}_{n,t}(\boldsymbol{k}) \rangle$ the Fermi velocity obtained by averaging over the Fermi surface, leading to

$$\text{Constant } \tau: \quad (\rho_0 \times \lambda)^{-1}_{\boldsymbol{t},\boldsymbol{\tau}} = \frac{e^2}{4\pi^3 \hbar} \langle \boldsymbol{v}_{n,t}(\boldsymbol{k}) \rangle \sum_n \iint_{S_F^n} \frac{\left|\boldsymbol{v}^2_{n,\boldsymbol{t}}(\boldsymbol{k})\right|}{|\boldsymbol{v}_n(\boldsymbol{k})|^2} dS, \tag{6}$$

Because of the well-known anisotropic conductivity of many MAX phases with low out-of-plane and high in-plane conductivity, only in-plane transport directions perpendicular to (001) have been considered in this work. Therefore, the assessment of the scalability of the conductivity of the MAX phases strictly addresses the in-plane conductivity only via $(\rho_0 \times \lambda)^{-1}_{\perp,\boldsymbol{\lambda}}$ and $(\rho_0 \times \lambda)^{-1}_{\perp,\boldsymbol{\tau}}$. This will be discussed further below. Note that in-plane transport in hexagonal systems, such as the MAX phases, is isotropic and does not depend on the choice of the in-plane transport direction [20]. This has explicitly been verified for several MAX phases.

In the following, we will use a combination of the cohesive energy and the $\rho_0 \times \lambda$ figure of merit (both assuming constant τ or constant λ) to assess the suitability of a broad range of MAX phases for interconnect applications, as applied to elemental metals before [14,17,24].

## III. Results and discussion

Using the above methodology, a total number of 170 potential MAX compounds have been assessed. The materials list in the screening process was based on experimentally reported MAX phases [31–35] with components M ∈ {Ti, V, Cr, Ta}, A ∈ {Al, Si, Ga, Ge, As, Cd, In}, and X ∈ {N, C}. For a given composition, the formation enthalpies of the different polymorphs were then calculated to identify thermodynamically stable phases. For most compositions, the enthalpy of formation was negative for both α- and β-polymorphs, which suggests that they may coexist [31–34,36]. The complete results of the screening are summarized in Tables 2 to 4 in Appendix A. Moreover, the dynamical stability of the MAX phases with negative formation enthalpy has been confirmed by the analysis of their phonon dispersion relations represented by stabilized phonon modes (positive frequencies) and the softening of acoustic modes near various high-symmetry points. By contrast, dynamical instabilities and structural transitions to different polymorphs or phases can be expected whenever an apparent phonon frequency becomes negative (imaginary) in some part of the dispersion relation.

In the following, we discuss in detail the case of M = V, A = Al and X = C, *i.e.* the $V_{n+1}AlC_n$ system, as an example. Table 1 shows the formation enthalpies of all polymorphs of the three considered $V_{n+1}AlC_n$ stoichiometries (211, 312, and 413). For the α-polymorphs, the calculated formation enthalpies $\Delta H_f$ were found to be almost identical when normalized to the number of atoms. In contrast, the β-polymorphs, which exist for the 312 and 413 phases, have higher (less negative) formation enthalpies. Nonetheless, the enthalpies of β-$V_3AlC_2$ and β-$V_4AlC_3$ are still negative, which suggests that they may coexist with the α-polymorphs, *e.g.* in thin films that are processed under conditions far from equilibrium. The dynamic stability of especially the β-polymorphs can be understood by the phonon dispersion relations in Fig. 2. Figure 2 indicates that the dispersion relations of all $V_{n+1}AlC_n$ structures show positive frequencies only (indicating dynamic stability) with the exception of β-$V_3AlC_2$, which exhibits negative phonon frequencies in some parts of the reciprocal space. This indicates that β-$V_3AlC_2$ is expected to be dynamically unstable and that a structural transition to another polymorph or phase may occur, despite a negative formation enthalpy. Note that this is not the case for β-$V_4AlC_3$, which is expected to be dynamically stable. Since the α- and β-polymorphs differ mainly by the stacking order of the hexagonal planes, this also implies that $V_4AlC_3$ may be strongly susceptible to stacking faults. As shown in Tables 2 to 4, such a behavior can be observed for several different MAX materials. From a standpoint of phase stability, the 211 phases thus appear generally preferable over the polymorphic 312 and 413 phases.

By contrast, the calculated cohesive energies per atom do not vary strongly with composition *n* and polymorph in the $V_{n+1}AlC_n$ system. The cohesive energy values are large, much larger than the cohesive energy per atom calculated for Cu within the same framework (3.8 eV), and comparable to that of Ru (8.0 eV). Since Ru has shown excellent resistance against EM without the need for liners as well as good dielectric reliability even in absence of diffusion barriers [3,43–45], it can be expected that the $V_{n+1}AlC_n$ MAX phases show a similar behavior, which is highly attractive for interconnect applications. The comprehensive results in Tables 2 to 4 show numerous C-based MAX materials (X = C) with cohesive energies per atom in this range. By contrast, N-based MAX materials (X = N) typically show lower cohesive energies for identical M and A atoms, although values still typically exceed the cohesive energy of Cu. Thus, MAX materials as such are promising to satisfy reliability requirements in scaled interconnects without the need for barrier or liner layers, with a clear potential advantage for C-based compounds.

The stoichiometry and the polymorphism have also a significant impact on the electronic structure and the morphology of the Fermi surface. Figure 3 depicts the Fermi surfaces of the different $V_{n+1}AlC_n$ phases. The color of the Fermi surfaces represents the corresponding Fermi velocities. As a reference, the Fermi surfaces of fcc Cu and hcp Ru are also shown. The figure shows that the morphology of the $V_{n+1}AlC_n$ Fermi surfaces and the typical group velocities differ strongly from those of Cu. While the Fermi surface of Cu is rather spherical and the Fermi velocity of Cu is nearly isotropic, the Fermi surfaces of the $V_{n+1}AlC_n$ MAX phases (and Ru) are strongly anisotropic. For Ru, this leads to different conductivities parallel and perpendicular to the hexagonal axis [46]. Fermi surfaces of MAX phases show even fewer electronic states along the hexagonal [001] axis due to their layered hexagonal structure, as shown in Fig. 3. This corresponds to a strong $\rho_0 \times \lambda$ anisotropy and leads to a severe anisotropy of the resistivity. The resistivity anisotropy has been experimentally confirmed for *e.g.* $V_2AlC$, $Cr_2AlC$, and $Ti_2AlC$ with ratios of out-of-plane and in-plane resistivities reaching several $10^3$ [36,47], although for other experimental studies, the anisotropy was markedly lower [34]. In this respect, MAX phases can be considered as 2D conductors and only the conductivity in the hexagonal plane is expected to be suitable for interconnect applications. For this reason, the $\rho_0 \times \lambda$ product was averaged over transport directions perpendicular to [001] only (denoted by the subscript $\perp$), as described in the previous section. All $\rho_0 \times \lambda$ values reported in this paper are thus valid for transport perpendicular to the hexagonal [001] axis.

The results for all studied MAX phases with negative formation enthalpies are listed in Tables 2 to 4 in Appendix A, including results for approximations of both constant isotropic τ or constant isotropic λ. In the following, we consider the most stable polymorphs only, *i.e.* the polymorphs with the lowest formation enthalpy for a given stoichiometry. In all cases, these polymorphs were dynamically stable with positive phonon frequencies only, as verified explicitly. To be technologically promising, a MAX compound should show a lower $\rho_0 \times \lambda$ than Cu ($(\rho_0 \times \lambda)_{\perp,\tau}$= $6.8\times10^{-16}$ Ωm²; $(\rho_0 \times \lambda)_{\perp,\lambda}$= $6.7\times10^{-16}$ Ωm²;) and, ideally, Ru ($(\rho_0 \times \lambda)_{\perp,\tau} = (\rho_0 \times \lambda)_{\perp,\lambda} = 5.1\times10^{-16}$ Ωm²). Tables 2 to 4 show that 80 out of the studied 170 MAX phases satisfy this criterion with both lower $(\rho_0 \times \lambda)_{\perp,\lambda}$ and $(\rho_0 \times \lambda)_{\perp,\tau}$ values than Cu as well as larger cohesive energy (> 3.8 eV). 69 out of the 80 compounds are correspond to the most stable phases with the lowest formation enthalpy for a given stoichiometry. Note that only 2 compounds have lower $(\rho_0 \times \lambda)_{\perp,\lambda}$ but higher $(\rho_0 \times \lambda)_{\perp,\tau}$ or *vice versa* with respect to Cu, which indicates that the nature of the approximation does not strongly affect the material selection process. When compared to Ru, 24 stable MAX compounds show similar cohesive energies (> 7 eV) and similar or lower $(\rho_0 \times \lambda)_{\perp,\lambda}$ and $(\rho_0 \times \lambda)_{\perp,\tau}$ (< 5.5×$10^{-16}$ Ω$m^2$). Again, the downselection does not depend on the approximation. Among the studied compounds with 211 stoichiometry, $V_2AlC$, $V_2SiC$, $V_2GaC$, $V_2GeC$, $V_2InC$, and $Ta_2GaC$ have the potential to perform similarly or better than Ru in terms of resistivity scalability (low $(\rho_0 \times \lambda)_{\perp,\lambda}$ and $(\rho_0 \times \lambda)_{\perp,\tau}$ products) and reliability (high cohesive energy). In addition, many N-based MAX phases show equally low $\rho_0 \times \lambda$ products below the Ru value, albeit with lower cohesive energies. Nonetheless, these compounds still show better $\rho_0 \times \lambda$ and $E_{coh}$ figure of merits than Cu and are therefore of potential interest as conductors in scaled interconnects. Analogously, promising stable compounds with 312 and 413 stoichiometries include α-$V_3SiC_2$, α-$V_3GaC_2$, α-$V_3GeC_2$, α-$V_3AsC_2$, α-$V_3CdC_2$, α-$V_3InC_2$, α-$V_3SnC_2$, α-$Ti_4AlC_3$, α-$Ti_4SiC_3$, α-$Ti_4GaC_3$, α-$Ti_4InC_3$, α-$V_4AlC_3$, α-$V_4GaC_3$, α-$V_4GeC_3$, α-$V_4CdC_3$, α-$V_4InC_3$, α-$V_4SnC_3$, and α-$Ta_4AlC_3$. Again, numerous N-based MAX phases show equally low $\rho_0 \times \lambda$ products as their C-based counterparts. However, their cohesive energy is again lower, rendering them less promising in terms of resistance to EM and the need for diffusion barriers. As for the 211 phases, a comparison with Cu is however generally more favorable and N-based MAX phases may still be considered as alternatives to Cu. All stable MAX phases (with the lowest formation enthalpy for a given stoichiometry) that show lower $\rho_0 \times \lambda$ products and higher cohesive energies than Cu are represented in the benchmark chart in Fig. 4.

The $V_{n+1}AlC_n$ system in Table 1 leads to insights that are confirmed more generally for other MAX phases also. First, the data do not support a clear trend with stoichiometry *n*. By looking at the full data set in Tables 2 to 4, no specific value for *n* can be identified as particularly promising. Hence, the choice of M, A, and X elements appears to dominate the effect of stoichiometry. Second, β-polymorphs have typically lower $\rho_0 \times \lambda$ products than α-polymorphs although there are several exceptions, as it can be seen in Tables 3 and 4. While some of them show negative phonon frequencies and may be dynamically unstable, the study and stabilization of metastable β-polymorphs may still be interesting for interconnect applications.

To further downselect promising MAX materials, the bulk resistivity needs to be considered. For several MAX materials, including e.g. $Cr_2AlC$ and $V_2AlC$, low bulk resistivities of the order of 10 μΩcm or below have been reported [32,36]. Such values lead together with low $\rho_0 \times \lambda$ values to MFPs of a few nm only, much smaller than the value of Cu (~40 nm) and comparable to Ru (~6 nm) [14,24]. By *ab initio* techniques, the accurate calculation of the bulk resistivities is very challenging and cannot be achieved for a broad range of materials. However, the current list can provide a first starting point for further experimental or theoretical study of MAX materials.

The anisotropic 2D conduction of MAX compounds leads to specific challenges for the integration in scaled interconnects. To realize scaled interconnect lines with low resistances, the transport needs to be perpendicular to the hexagonal [001] axis, which necessitates MAX materials with (001) texture. This indicates that the control of the grain orientation during the deposition process will be key for interconnect applications. Such texture has been commonly observed for thin films of hcp elemental metals, *e.g.* Ru [24,48], and has also been achieved for $Cr_2AlC$, $Ti_2AlC$, and $Ti_3AlC_2$ thin films [48]; in other cases, suitable templates, such as TiC, were employed [50,51]. Future work will be needed to demonstrate that composition, phase, polymorph, and texture can be controlled sufficiently for interconnect applications, especially at temperatures below 450°C, compatible with back-end-of-line CMOS processing. Finally, we remark that the anisotropic 2D conduction of the MAX compounds may also have advantages, since the strong anisotropy of the Fermi surface is expected to suppress surface scattering at (001) surfaces [41], reducing even further the sensitivity to the dimension of scaled interconnects.

## IV. CONCLUSIONS

In this work, we have reported on the potential of ternary MAX compounds as conductor materials in scaled interconnects, with the goal to replace Cu used in present CMOS logic technology. Automated first-principles calculations have been used to identify thermodynamically stable MAX phases. Further benchmarking of MAX phases was performed using the cohesive energy as a proxy for reliability and the need for diffusion barriers as well as the product of the resistivity and the MFP of the charge carriers, assuming either a constant relaxation time τ, $(\rho_0 \times \lambda)_{\perp,\tau}$, or a constant MFP λ $(\rho_0 \times \lambda)_{\perp,\lambda}$, as proxies for the resistivity scaling at small dimensions. Among the studied 170 MAX compounds, 69 stable phases showed a higher cohesive energy and a lower $\rho_0 \times \lambda$ product than Cu, which renders them of potential interest for interconnect metallization applications. 24 of those stable compounds still compared favorably to Ru within the same benchmarking framework. Our findings suggest that MAX phases can be potential contenders to replace Cu in scaled interconnects in future technology nodes. Open issues include the integration of highly (001) textured films in interconnects, which is necessary due to the strong conductivity anisotropy of the MAX phases, as well as deposition at low temperatures that are compatible with back-end-of-line processing.

### ACKNOWLEDGEMENTS

This work has been supported by imec's industrial affiliate program on nano-interconnects.

### APPENDIX A: PROPERTIES OF ALL THERMODYNAMICALLY STABLE MAX PHASES

The properties (formation enthalpy $\Delta H_f$, cohesive energy $E_{coh}$, $(\rho_0 \times \lambda)_{\perp,\lambda}$, and $(\rho_0 \times \lambda)_{\perp,\tau}$) of all studied thermodynamically stable ($\Delta H_f < 0$) MAX phases obtained from the ab initio calculations are listed in Tab. 2 to 4. Table 2 contains results for phases with 211 stoichiometry, Table 3 for phases with 312 stoichiometry, and Table 4 for phases with 413 stoichiometry.

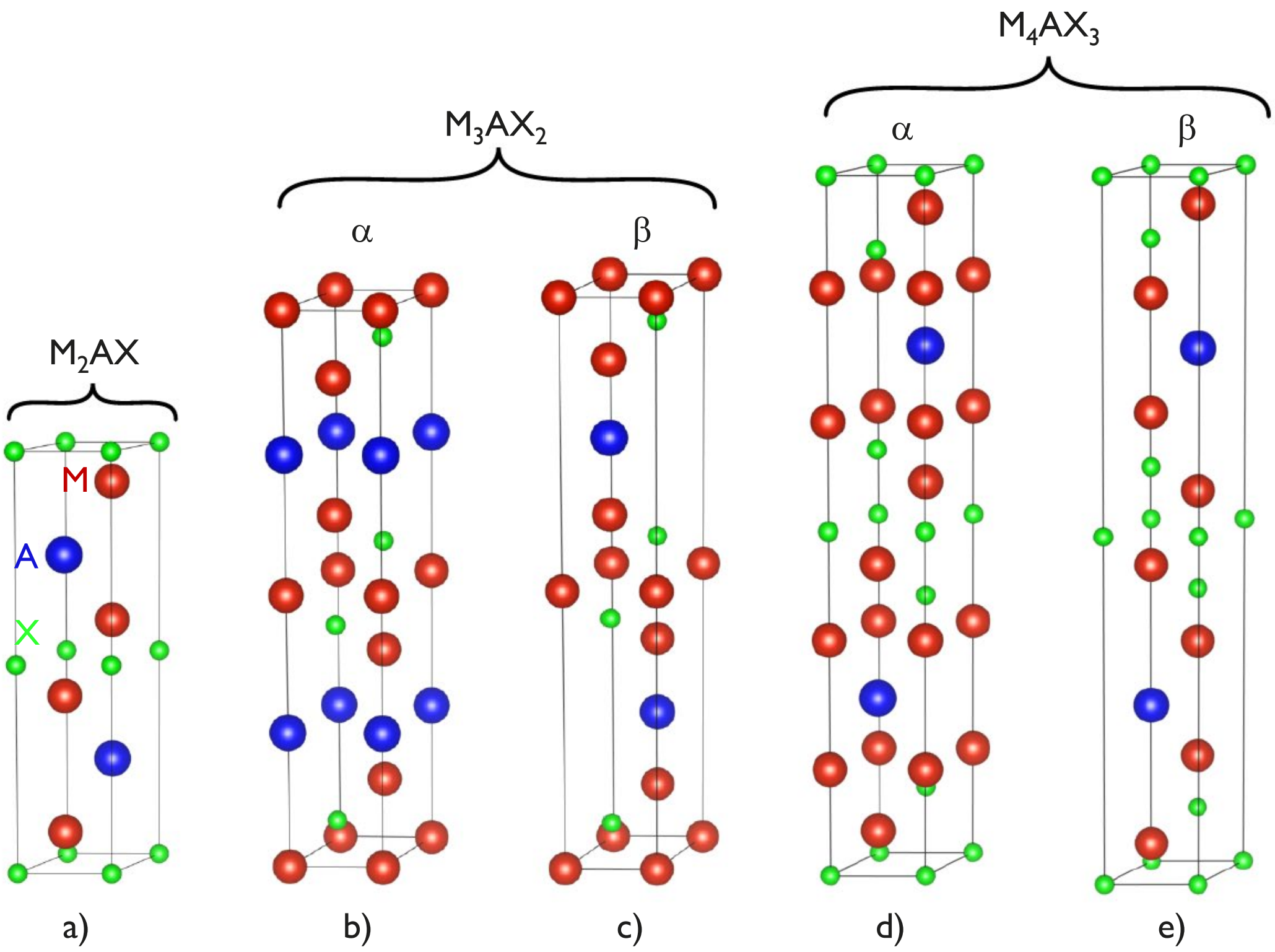

**Figure 1.** Schematic illustration of the crystal structures of the $M_{n+1}AX_n$ phase prototypes, including their polymorphs. Here, *n* is taking values of 1 (a), 2 (b and c), and 3 (d and e). Red dots represent the M atoms, blue dots the A atoms, and green dots the X atoms.

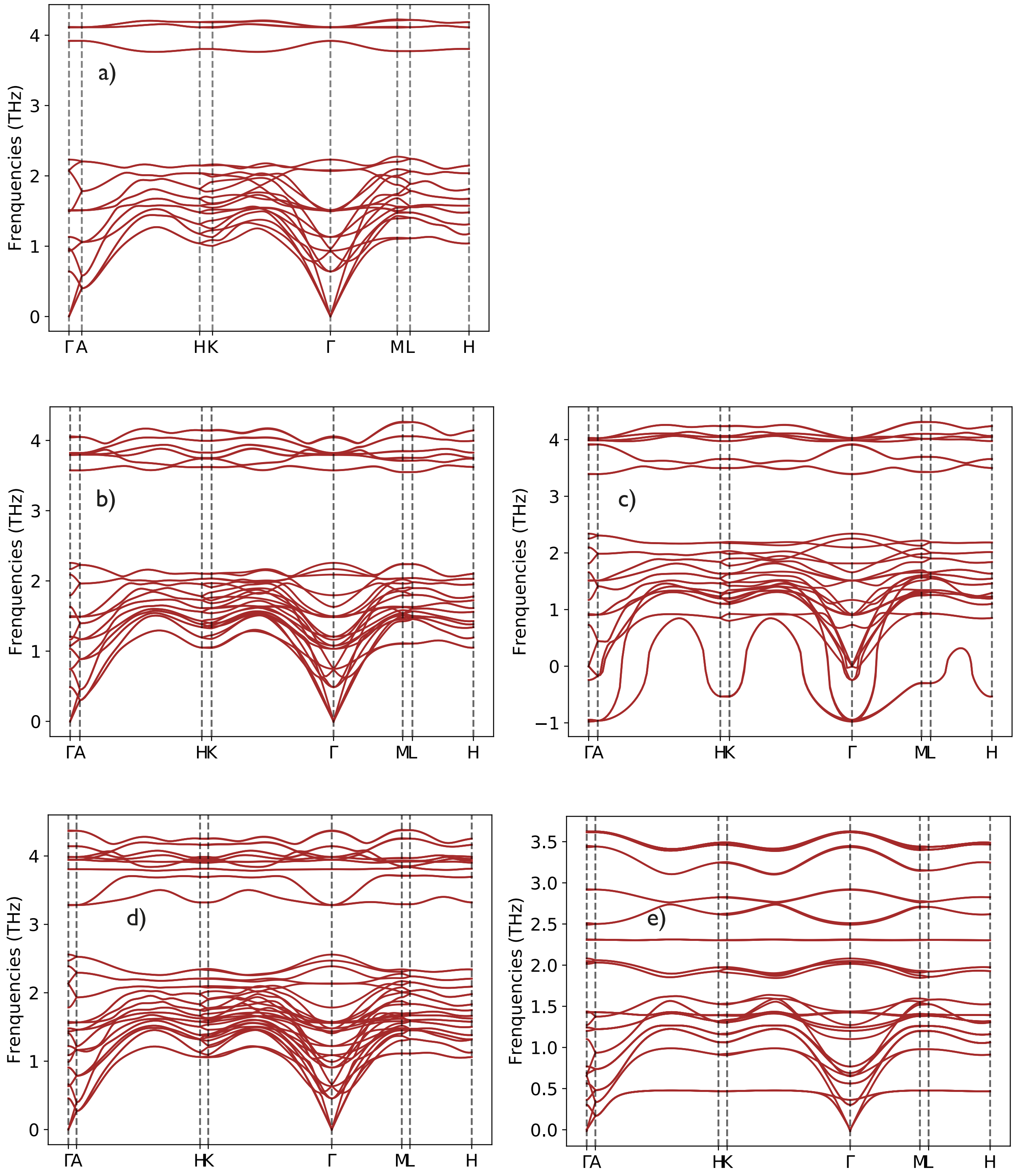

**Figure 2.** Calculated phonon dispersion curves of the $V_{n+1}AlC_n$ MAX phases: (a) $V_2AlC$, (b) α-$V_3AlC_2$, (c) β-$V_3AlC_2$, (d) α-$V_4AlC_3$, and (e) β-$V_4AlC_3$. The dispersion relations of all phases contain only positive frequencies, indicating that they dynamically stable, except for the β-$V_3AlC_2$ polymorph, which shows also negative phonon frequencies and is expected to show dynamic instabilities and eventually undergo a structural transition.

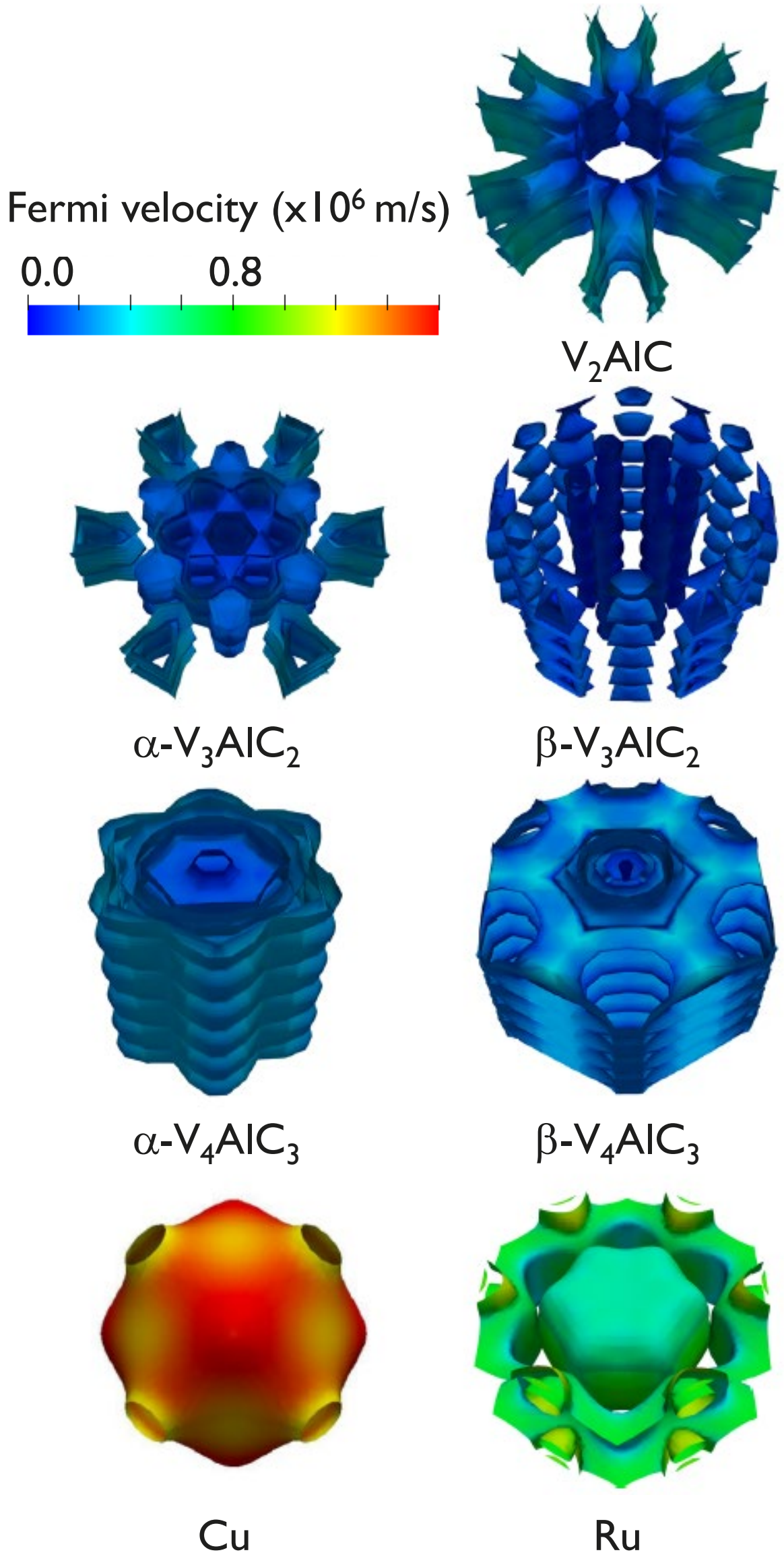

**Figure 3.** Calculated Fermi surfaces of the $V_{n+1}AlC_n$ MAX phases, as well as Cu and Ru as references. The surface color represents the Fermi velocity.

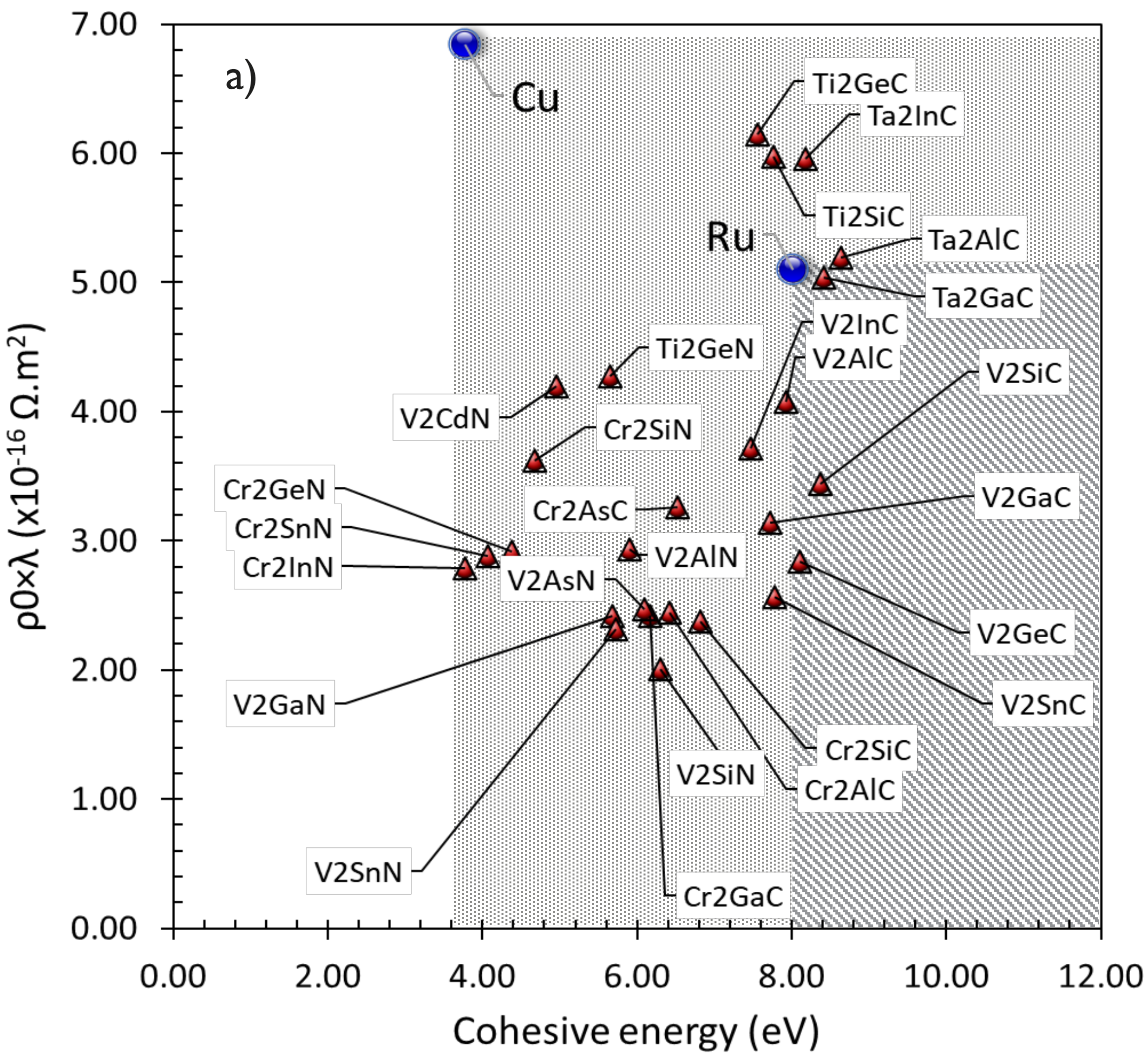

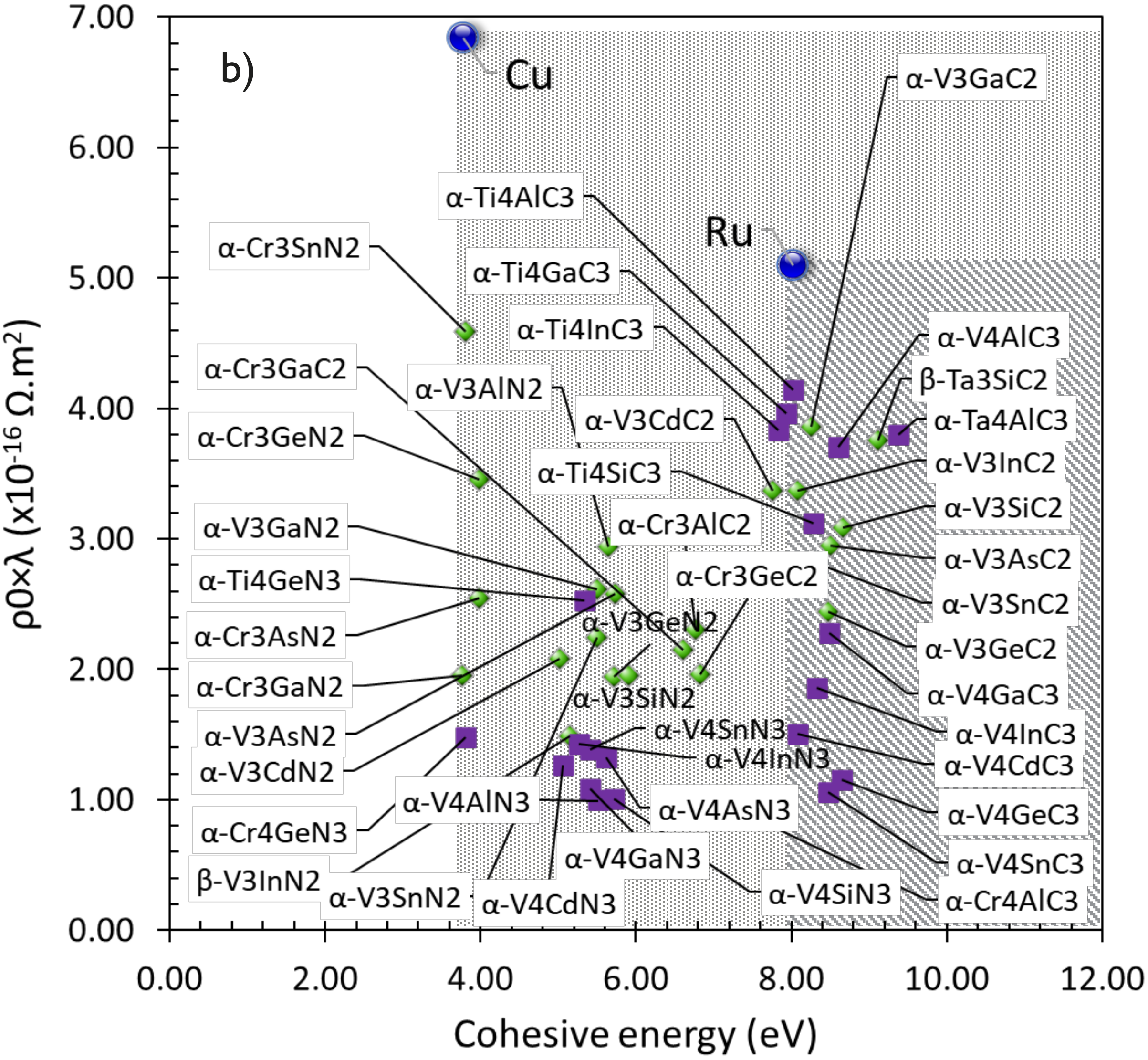

**Figure 4.** Figure of merits for stable MAX phases with respect to Cu and Ru as reference materials: resistivity scalability potential ($\rho_0 \times \lambda$) and resistance to EM (cohesive energy) for the 211 stoichiometry (left) and the 312 and 413 stoichiometries (right). The light and dark grey areas represent the regions, where MAX phases may be expected to have favorable properties with respect to Cu and Ru, respectively. For clarity, MAX phases outside the grey areas are not included.

**Table I.** Computed enthalpies of formation ($\Delta H_f$) of the different crystal polymorphs (α and β) of the $V_{n+1}AlC_n$ MAX phases (M = V, A = Al, X = C) together with their cohesive energy $E_{coh}$ and the $\rho_0 \times \lambda$ product. The lowest formation enthalpy corresponds to the most thermodynamically favorable structure at a given stoichiometry. The different stoichiometries are referred to as 211, 312 and 413. Note that the 211 stoichiometry has no polymorphs. For comparison: constant τ, $(\rho_0 \times \lambda)_{\perp,\tau}$ = 6.8×10$^{-16}$ Ωm$^2$ and 5.1×10$^{-16}$ Ωm$^2$ for Cu and Ru, respectively; constant $\lambda$, $(\rho_0 \times \lambda)_{\perp,\lambda}$ = 6.7×10$^{-16}$ Ωm$^2$ and 5.1×10$^{-16}$ Ωm$^2$ for Cu and Ru, respectively [14].

| $V_{n+1}AlC_n$ | | $n$ = 1 (211) | $n$ = 2 (312) | $n$ = 3 (413) |
|---|---|---|---|---|
| α | $\Delta H_f$ (eV/atom) | −0.528 | −0.531 | −0.504 |
| | $E_{coh}$ (eV/atom) | 7.9 | 8.4 | 8.6 |
| | $(\rho_0 \times \lambda)_{\perp,\tau}$ (10$^{-16}$ Ωm$^2$) | 4.1 | 7.5 | 3.7 |
| | $(\rho_0 \times \lambda)_{\perp,\lambda}$ (10$^{-16}$ Ωm$^2$) | 5.3 | 7.3 | 3.7 |
| β | $\Delta H_f$ (eV) | | −0.260 | −0.253 |
| | $E_{coh}$ (eV/atom) | | 8.1 | 7.8 |
| | $(\rho_0 \times \lambda)_{\perp,\tau}$ (10$^{-16}$ Ωm$^2$) | | 4.8 | 3.2 |
| | $(\rho_0 \times \lambda)_{\perp,\lambda}$ (10$^{-16}$ Ωm$^2$) | | 5.4 | 3.8 |

**Table 2:** Calculated formation enthalpy $\Delta H_f$, cohesive energy $E_{coh}$, as well as $\rho_0 \times \lambda$ products for MAX phases with $\Delta H_f < 0$ and 211 stoichiometry ($M_2AX$). Values of $\rho_0 \times \lambda$ are listed for in-plane transport as well as both under assumptions of constant isotropic relaxation time $\tau$ ($(\rho_0 \times \lambda)_{\perp,\tau}$, Eq. (6)) or constant isotropic mean free path $\lambda$ ($(\rho_0 \times \lambda)_{\perp,\lambda}$, Eq. (4)).

| **Material** | **$\Delta H_f$ (eV/atom)** | **$E_{coh}$ (eV/atom)** | **$(\rho_0 \times \lambda)_{\perp,\tau}$ ($10^{-16}$ Ωm²)** | **$(\rho_0 \times \lambda)_{\perp,\lambda}$ ($10^{-16}$ Ωm²)** |
|---|---|---|---|---|
| $Ti_2AlC$ | −3.110 | 7.3 | 7.5 | 7.6 |
| $Ti_2AlN$ | −3.148 | 5.4 | 7.6 | 6.5 |
| $Ti_2SiC$ | −0.805 | 7.8 | 6.0 | 6.1 |
| $Ti_2GaN$ | −0.823 | 5.2 | 10.1 | 10.6 |
| $Ti_2GeC$ | −0.687 | 7.6 | 6.1 | 6.1 |
| $Ti_2GeN$ | −3.136 | 5.6 | 4.3 | 4.2 |
| $Ti_2InC$ | −3.010 | 6.9 | 9.2 | 8.2 |
| $Ti_2InN$ | −0.780 | 5.0 | 13.1 | 13.0 |
| $V_2AlC$ | −0.528 | 7.9 | 4.1 | 5.3 |
| $V_2AlN$ | −2.727 | 5.9 | 2.9 | 3.5 |
| $V_2SiC$ | −0.545 | 8.4 | 3.4 | 2.7 |
| $V_2SiN$ | −2.706 | 6.3 | 2.0 | 3.3 |
| $V_2GaC$ | −0.532 | 7.7 | 3.1 | 5.2 |
| $V_2GaN$ | −2.709 | 5.7 | 2.4 | 4.4 |
| $V_2GeC$ | −0.496 | 8.1 | 2.8 | 3.6 |
| $V_2AsN$ | −2.676 | 6.1 | 2.5 | 4.2 |
| $V_2CdC$ | −0.311 | 7.0 | 8.1 | 12.4 |
| $V_2CdN$ | −2.520 | 5.0 | 4.2 | 4.5 |
| $V_2InC$ | −0.361 | 7.5 | 3.7 | 3.9 |
| $V_2SnN$ | −2.527 | 5.7 | 2.3 | 4.3 |
| $Cr_2AlC$ | −0.188 | 6.4 | 2.4 | 3.4 |
| $Cr_2SiC$ | −0.155 | 6.8 | 2.4 | 2.8 |
| $Cr_2SiN$ | −2.240 | 4.7 | 3.6 | 4.0 |
| $Cr_2GaC$ | −0.148 | 6.2 | 2.4 | 3.6 |
| $Cr_2GeN$ | −2.174 | 4.4 | 2.9 | 2.8 |
| $Cr_2AsC$ | −0.059 | 6.5 | 3.3 | 3.3 |
| $Cr_2AsN$ | −2.153 | 4.4 | 11.5 | 12.1 |
| $Cr_2InN$ | −2.058 | 3.8 | 2.8 | 3.1 |
| $Cr_2SnN$ | −2.051 | 4.1 | 2.9 | 5.6 |
| $Ta_2AlC$ | −0.535 | 8.6 | 5.2 | 6.3 |
| $Ta_2SiC$ | −0.532 | 9.1 | 13.4 | 13.4 |
| $Ta_2GaC$ | −0.526 | 8.4 | 5.0 | 5.4 |
| $Ta_2InC$ | −0.383 | 8.2 | 6.0 | 7.2 |

**Table 3:** Calculated formation enthalpy $\Delta H_f$, cohesive energy $E_{coh}$, as well as $\rho_0 \times \lambda$ products for MAX phases with $\Delta H_f < 0$ and 312 stoichiometry ($M_3AX_2$). Values of $\rho_0 \times \lambda$ are listed for in-plane transport as well as both under assumptions of constant isotropic relaxation time $\tau$ ($(\rho_0 \times \lambda)_{\perp,\tau}$, Eq. (6)) or constant isotropic mean free path $\lambda$ ($(\rho_0 \times \lambda)_{\perp,\lambda}$, Eq. (4)).

| Material | $\Delta H_f$ (eV/atom) | $E_{coh}$ (eV/atom) | $(\rho_0 \times \lambda)_{\perp,\tau}$ ($10^{-16}$ Ωm²) | $(\rho_0 \times \lambda)_{\perp,\lambda}$ ($10^{-16}$ Ωm²) |
|---|---|---|---|---|
| α-$Ti_3AlC_2$ | −0.780 | 7.8 | 10.4 | 10.0 |
| α-$Ti_3SiC_2$ | −0.838 | 8.1 | 7.3 | 7.7 |
| β-$Ti_3SiC_2$ | −0.507 | 7.8 | 2.4 | 2.6 |
| α-$Ti_3GaN_2$ | −3.954 | 5.2 | 8.4 | 7.9 |
| β-$Ti_3GaN_2$ | −3.689 | 4.9 | 3.4 | 3.9 |
| α-$Ti_3GeC_2$ | −0.856 | 8.0 | 7.6 | 8.0 |
| α-$Ti_3GeN_2$ | −3.946 | 5.4 | 9.8 | 8.6 |
| β-$Ti_3GeN_2$ | −3.701 | 5.2 | 7.7 | 7.1 |
| α-$Ti_3InC_2$ | −0.763 | 7.6 | 10.0 | 8.1 |
| α-$Ti_3InN_2$ | −3.860 | 5.0 | 9.0 | 9.4 |
| β-$Ti_3InN_2$ | −3.637 | 4.8 | 2.9 | 2.8 |
| α-$Ti_3SnC_2$ | −0.802 | 7.8 | 8.1 | 8.2 |
| β-$Ti_3SnC_2$ | −0.574 | 7.6 | 2.6 | 2.6 |
| α-$V_3AlC_2$ | −0.531 | 8.4 | 7.5 | 7.3 |
| β-$V_3AlC_2$ | −0.260 | 8.1 | 4.8 | 5.4 |
| α-$V_3AlN_2$ | −3.415 | 5.7 | 2.9 | 3.5 |
| α-$V_3SiC_2$ | −0.508 | 8.7 | 3.1 | 4.0 |
| α-$V_3SiN_2$ | −3.392 | 5.9 | 2.0 | 2.3 |
| α-$V_3GaC_2$ | −0.528 | 8.3 | 3.9 | 4.1 |
| α-$V_3GaN_2$ | −3.411 | 5.5 | 2.6 | 2.9 |
| α-$V_3GeC_2$ | −0.478 | 8.5 | 2.4 | 3.5 |
| α-$V_3GeN_2$ | −3.359 | 5.7 | 2.0 | 3.4 |
| α-$V_3AsC_2$ | −0.480 | 8.5 | 3.0 | 2.9 |
| α-$V_3AsN_2$ | −3.351 | 5.7 | 2.6 | 2.8 |
| α-$V_3CdC_2$ | −0.384 | 7.8 | 3.4 | 3.7 |
| α-$V_3CdN_2$ | −3.287 | 5.0 | 2.1 | 2.1 |
| α-$V_3InC_2$ | −0.413 | 8.1 | 3.4 | 3.9 |
| β-$V_3InN_2$ | −3.121 | 5.2 | 1.5 | 2.7 |
| α-$V_3SnC_2$ | −0.380 | 8.3 | 2.8 | 4.6 |
| α-$V_3SnN_2$ | −3.255 | 5.5 | 2.3 | 3.8 |
| α-$Ta_3AlC_2$ | −0.248 | 9.1 | 135 | 137 |
| α-$Cr_3AlC_2$ | −0.072 | 6.8 | 2.3 | 2.6 |
| α-$Cr_3GaC_2$ | −0.052 | 6.6 | 2.2 | 2.2 |
| α-$Cr_3GaN_2$ | −2.838 | 3.8 | 2.0 | 2.5 |
| α-$Cr_3GeC_2$ | −0.003 | 6.8 | 2.0 | 2.8 |
| α-$Cr_3GeN_2$ | −2.793 | 4.0 | 3.5 | 3.9 |
| α-$Cr_3AsN_2$ | −2.769 | 4.0 | 2.6 | 3.0 |
| α-$Cr_3CdN_2$ | −2.730 | 3.3 | 2.7 | 3.2 |
| α-$Cr_3InN_2$ | −2.744 | 3.6 | 2.5 | 3.0 |
| α-$Cr_3SnN_2$ | −2.728 | 3.8 | 4.6 | 6.6 |
| β-$Ta_3AlC_2$ | −3.234 | 8.8 | 14.6 | 15.7 |
| α-$Ta_3AlN_2$ | −0.528 | 6.2 | 6.0 | 5.6 |
| β-$Ta_3AlN_2$ | −2.959 | 5.9 | 11.0 | 11.2 |
| α-$Ta_3SiC_2$ | −0.272 | 9.4 | 9.8 | 8.2 |
| β-$Ta_3SiC_2$ | −0.582 | 9.1 | 3.8 | 3.9 |

**Table 4:** Calculated formation enthalpy $\Delta H_f$, cohesive energy $E_{coh}$, as well as $\rho_0 \times \lambda$ products for MAX phases with $\Delta H_f < 0$ and 413 stoichiometry ($M_4AX_3$). Values of $\rho_0 \times \lambda$ are listed for in-plane transport as well as both under assumptions of constant isotropic relaxation time $\tau$ ($(\rho_0 \times \lambda)_{\perp,\tau}$, Eq. (6)) or constant isotropic mean free path $\lambda$ ($(\rho_0 \times \lambda)_{\perp,\lambda}$, Eq. (4)).

| **Material** | **$\Delta H_f$ (eV/atom)** | **$E_{coh}$ (eV/atom)** | **$(\rho_0 \times \lambda)_{\perp,\tau}$ ($10^{-16}$ Ωm$^2$)** | **$(\rho_0 \times \lambda)_{\perp,\lambda}$ ($10^{-16}$ Ωm$^2$)** |
|---|---|---|---|---|
| α-$Ti_4AlC_3$ | −0.792 | 8.0 | 4.1 | 4.3 |
| α-$Ti_4SiC_3$ | −0.835 | 8.3 | 3.1 | 3.2 |
| β-$Ti_4SiC_3$ | −0.070 | 7.5 | 2.9 | 2.0 |
| α-$Ti_4GaC_3$ | −0.811 | 7.9 | 4.0 | 4.1 |
| β-$Ti_4GaC_3$ | −0.005 | 7.1 | 2.4 | 2.4 |
| α-$Ti_4GeN_3$ | −4.346 | 5.3 | 2.5 | 2.8 |
| β-$Ti_4GeN_3$ | −3.895 | 4.9 | 3.1 | 3.2 |
| α-$Ti_4InC_3$ | −0.744 | 7.8 | 3.8 | 3.8 |
| β-$Ti_4InN_3$ | −3.749 | 4.5 | 0.6 | 2.2 |
| α-$V_4AlC_3$ | −0.504 | 8.6 | 3.7 | 3.7 |
| β-$V_4AlC_3$ | −0.253 | 7.8 | 3.2 | 3.8 |
| α-$V_4AlN_3$ | −3.759 | 5.5 | 1.0 | 1.3 |
| β-$V_4AlN_3$ | −3.329 | 5.1 | 3.2 | 3.5 |
| α-$V_4SiN_3$ | −3.753 | 5.7 | 1.0 | 1.1 |
| α-$V_4GaC_3$ | −0.502 | 8.5 | 2.2 | 2.8 |
| α-$V_4GaN_3$ | −3.762 | 5.4 | 1.1 | 1.4 |
| α-$V_4GeC_3$ | −0.460 | 8.7 | 1.1 | 1.2 |
| α-$V_4AsN_3$ | −3.735 | 5.6 | 1.3 | 1.5 |
| α-$V_4CdC_3$ | −0.357 | 8.1 | 1.5 | 1.6 |
| α-$V_4CdN_3$ | −3.671 | 5.1 | 1.3 | 1.7 |
| α-$V_4InC_3$ | −0.384 | 8.3 | 1.9 | 1.9 |
| α-$V_4InN_3$ | −3.661 | 5.3 | 1.4 | 2.2 |
| α-$V_4SnC_3$ | −0.365 | 8.5 | 1.1 | 1.2 |
| α-$V_4SnN_3$ | −3.649 | 5.4 | 1.4 | 1.8 |
| α-$Cr_4AlC_3$ | −0.010 | 6.9 | 1.0 | 1.1 |
| β-$Cr_4SiN_3$ | −2.672 | 3.5 | 2.9 | 4.3 |
| α-$Cr_4GeN_3$ | −3.125 | 3.8 | 1.5 | 1.4 |
| α-$Ta_4AlC_3$ | −0.582 | 9.4 | 3.8 | 3.3 |
| β-$Ta_4SiC_3$ | −0.717 | 8.5 | 26.0 | 26.4 |